\documentclass[11pt,twoside]{article}

\usepackage{asp2006}
\usepackage{epsf}
\usepackage{psfig}
\usepackage{lscape}

\def\ergs{$~\mathrm{ergs}~\mathrm{s}^{-1}$}

\begin{document}
\title{H.E.S.S. observations of massive stellar clusters}   
\author{Stefan Ohm\altaffilmark{1},
  Dieter Horns\altaffilmark{2},
  Olaf Reimer\altaffilmark{3,4},
  Jim Hinton\altaffilmark{5},
  Gavin Rowell\altaffilmark{6}, 
  Emma de O\~na Wilhelmi\altaffilmark{1}, 
  Milton Virgilio Fernandes\altaffilmark{2}, 
  Fabio Acero\altaffilmark{7}, 
  Alexandre Marcowith\altaffilmark{7}, 
  for the H.E.S.S. Collaboration\altaffilmark{8}}   

\altaffiltext{1}{Max-Planck-Institut f\"ur Kernphysik, P.O. Box 103980, 69029 Heidelberg, Germany}
\altaffiltext{2}{Universit\"at Hamburg, Institut f\"ur Experimentalphysik, Luruper Chaussee 149, 
22761 Hamburg, Germany}
\altaffiltext{3}{Stanford University, HEPL \& KIPAC, Stanford, CA 94305-4085, USA}
\altaffiltext{4}{Universit\"at Innsbruck, Institut f\"ur Astro- und Teilchenphysik, Technikerstrasse 
25/8, 6020 Innsbruck, Austria}
\altaffiltext{5}{School of Physics \& Astronomy, University of Leeds, Leeds LS2 9JT, UK}
\altaffiltext{6}{School of Chemistry \& Physics, University of Adelaide, Adelaide 5005, Australia}
\altaffiltext{7}{Laboratoire de Physique th´eorique et astroparticules, universit´e Montpellier II, 
CNRS/IN2P3, Montpellier, France}
\altaffiltext{8}{http://www.mpi-hd.mpg.de/hfm/HESS/}    

\begin{abstract} 
Stellar clusters are potential acceleration sites of very-high-energy (VHE, E$>$ 100GeV) 
particles since they host supernova remnants (SNRs) and pulsar wind nebulae (PWNe). Additionally, 
in stellar clusters, particles can also be accelerated e.g. at the boundaries of wind-blown 
bubbles, in colliding wind zones in massive binary systems or in the framework of collective wind 
or wind/supernova(SN) ejecta scenarios. Motivated by the detection of VHE 
$\gamma$-ray emission towards Westerlund 2 and assuming similar particle acceleration mechanisms at 
work, Westerlund 1 is an even more promising target for VHE $\gamma$-ray observations given that 
massive star content and distance are more favorable for detectable VHE $\gamma$-ray emission 
compared to Westerlund 2. Here, H.E.S.S. observations of massive stellar clusters 
in general with special emphasis on the most massive stellar cluster in the galaxy, Westerlund 1 
are summarized.
\end{abstract}



\section{Introduction}

In recent years the field of ground-based VHE $\gamma$-ray astronomy experienced a 
scientific breakthrough due to developments in instrumentation, operation and data 
analysis of the 3$^{rd}$ generation of Imaging Atmospheric Cherenkov Telescopes (IACTs). 
Telescope systems like H.E.S.S. \citep{Hinton2004}, MAGIC \citep{Lorenz2004}, VERITAS 
\citep{Weekes2002} or CANGAROO-III \citep{Kubo2004} opened a previously inaccessible 
window for the study of astrophysical objects at very high energies. Especially the 
increase of the number of detected VHE $\gamma$-ray sources by one order of magnitude 
is a merit of the Galactic Plane Scan (GPS) performed by the H.E.S.S. Collaboration 
between 2004 and 2008 \citep{Aharonian2006,Hoppe2007,Chaves2008}. 
In the GPS, which covers basically the whole inner galaxy between Galactic 
longitude l $\sim$ 275$^\circ$ and $\sim$ 60$^\circ$, a rich diversity of astrophysical 
objects which emit VHE $\gamma$-rays was discovered. Even if a large part of the source 
population lacks of a firm detection in other wavelength bands, a 
significant fraction of identified objects are connected to the late phases of stellar 
evolution like e.g. supernova remnants (SNRs) or pulsar wind nebulae (PWN). Moreover, one 
class of objects emitting VHE $\gamma$-rays could be associated to regions of massive 
star formation and to the birthplaces of the massive progenitors of some SNRs and PWNe, 
massive stellar clusters. 

In this work H.E.S.S. observations of massive stellar clusters with special emphasize 
on the most massive stellar cluster in our galaxy, Westerlund 1 are presented. After an 
introduction of the acceleration mechanisms which are at work in such systems (Chapter 
\ref{section:theory}), recent results of multiwavelength observations and the possible 
connection of the unidentified source HESS~J1614--518 to the stellar cluster Pismis 22 
are presented in Chapter \ref{section:Pismis22} The detection of Westerlund 2 
(HESS~J1023--575) in VHE $\gamma$-rays is discussed in Chapter \ref{section:Wd2} 
Finally, the detection of VHE $\gamma$-ray emission from the vicinity of the massive 
stellar cluster Westerlund 1 is discussed in Chapter \ref{section:Wd1}

\section{Particle acceleration mechanisms in massive stellar clusters} \label{section:theory}
It is widely believed that massive stars form in groups from the collapse of gas 
condensations in clumps, inside giant molecular clouds (e.g. \citet{Heyer1998,Williams2000,
Dame2001,Jackson2006}). The accretion of cloud material onto the massive stars is stopped, when 
their winds, outflows and UV radiation leads to the dissipation or disruption of the natal 
molecular cloud. Depending on the total mass of the system, these groups of stars end as loosely 
bound {\it associations} or as dense gravitationally bound {\it stellar clusters}.

\subsection{Massive binary systems}
Since a large fraction of massive stars occur in binary (or even triple, quadruple, etc.) 
systems (e.g. \citep{Zinnecker2003,Gies2008} and references therein), the collision of their 
strong supersonic winds with terminal velocities v$_\infty> 1000-5000$~km~s$^{-1}$ 
\citep{Cassinelli1979}, produces strong shocks. Through first-order Fermi acceleration electrons 
and protons can be accelerated to high energies \citep{Eichler1993}. \citet{Muecke2002},
\citet{Benaglia2003} and \citet{Reimer2006} showed, that detectable $\gamma$-radiation up to GeV 
energies can be generated through inverse Compton scattering of relativistic electrons in the dense 
photospheric stellar radiation field in the wind-wind collision zone. The detection of non-thermal 
X-ray emission in massive binary systems like e.g. WR140 \citep{Dougherty2005} has proven that 
indeed electrons are accelerated in the wind collision region to relativistic energies.
However, just in one hadronic emission scenario (proposed by 
\citet{Bednarek2005}), a detectable signal by current and future IACTs is predicted. Here, the 
inelastic scattering of relativistic nucleons on particles in the dense stellar wind produces a 
significant amount of neutral pions which then decay into VHE $\gamma$-rays. Though, at these high 
energies, $\gamma$-rays suffer from $\gamma\gamma$~absorption which will diminish the observable 
flux from a close binary system.

\subsection{Collective stellar winds}
Given the fact, that massive stars mostly occur in associations and stellar clusters, the interaction 
of their strong supersonic winds leads to the creation of a large collective bubble, also called 
superbubble (SB), which is filled with a hot and tenuous plasma (e.g. \citep{Weaver1977}). Turbulent 
particle acceleration can then take place, where the interaction of wind material forms regions of strong 
turbulence and MHD waves (e.g. \citep{Byk01,Parizot2004,Higdon2005,Dwarkadas2008}).

\subsection{SN explosions in stellar clusters}
Since massive stars ($M \geq 8 M_{\odot}$) end their lives as SN explosions after a few Myrs, 
an additional contribution of kinetic energy is available to accelerate particles to very high energies.
The SNR shell will grow quicker due to the lower density in a SB and in a medium with higher sound 
speed, due to the higher temperature. This may result in efficient particle acceleration at MHD 
turbulences and at the boundary of the SB \citep{Parizot2004,Tang2005,Ferrand2009}.

\subsection{Implications}
In a simplified picture, the aforementioned acceleration mechanisms infer basic properties of 
stellar clusters, which in principle could indicate the fraction of total kinetic energy in 
non-thermal particles and therefore a possible emission in VHE $\gamma$-rays. One obvious 
property is the total mass of the cluster, since this directly determines the number of massive 
stars and therefore the total available mechanical energy in stellar winds (and later in the 
expanding shells of SN explosions). The age of the stellar cluster (and its mass) further implies 
whether or not the most massive members already evolved into SNe and finally, the fraction of 
massive stars located in binary systems defines the contribution from colliding wind binaries (CWBs).

In the next Chapters H.E.S.S. detections of VHE $\gamma$-ray sources which could 
be associated to stellar clusters of different mass, age and distance will be presented.

\section{HESS~J1614--518} \label{section:Pismis22}
\begin{figure}
\centering{\plotone{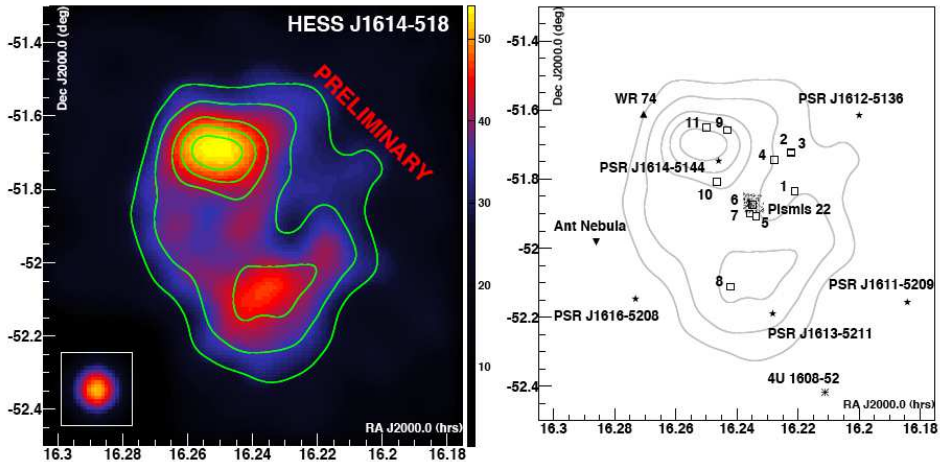}}
\caption{Left: Exposure corrected skymap of the VHE $\gamma$-ray excess, smoothed with a
Gaussian ($\sigma$ = 4.2'). The green contours represent significance levels of 3, 5, 7, 8, and
9$\sigma$, after integrating events within an oversampling radius of $\theta$ =0.1$^\circ$. 
The bottom left inlay shows how a point-like source would have been seen by H.E.S.S. for this 
analysis after Gaussian smoothing. Right: Finder chart for various objects discussed in the text. 
Black unfilled squares depict X-ray sources. Member and surrounding field stars are also shown 
for the open cluster Pismis 22 (from \citet{Piatti2000}, labeled as Pismis 22). Figure taken from 
\citet{Rowell2008}.
}\label{fig:HESSJ1614}
\end{figure}

HESS~J1614--518 is one of the brightest ($\sim25\%$ Crab flux, $E>200$~GeV) unidentified 
VHE $\gamma$-ray sources discovered during the Galactic plane scan~\citep{Aharonian2006}. 
The region was observed for a total of $\sim 13$ hours resulting in the detection of an 
extended, elliptical shaped emission with a half-width of $\sim0.4^\circ$ along the semimajor 
axis (Fig. \ref{fig:HESSJ1614}, left). The overall morphology of the source is well described by 
a double peak Gaussian profile ($\chi^2=0.7$). The differential energy 
spectrum is compatible with a power law (dN/dE$= \Phi_0 \cdot (\mbox{E}/1\,\mbox{TeV})^{- \Gamma}$) 
with a photon index of $\Gamma=2.26 \pm 0.05_{\mathrm{stat}} \pm 0.06_{\mathrm{syst}}$ and a 
normalization at 1\,TeV of $\Phi_0 = (7.83 \pm 0.40_{\mathrm{stat}} \pm 0.80_{\mathrm{syst}}) 
\times 10^{-12}$\,TeV$^{-1}$\,cm$^{-2}$\,s$^{-1}$.

An analysis of the MSX IR data shows several HII regions and molecular cloud complexes 
flanking the observed VHE $\gamma$-ray emission~\citep{Amaral1991,Matsunaga2001,Russeil2005,Li2006}. 
A search for potential counterparts reveals several interesting candidates (see Fig. 
\ref{fig:HESSJ1614}, right). On the basis of spin-down power of the pulsars in the field-of-view of 
HESS~J1614--518 none of these is powerful enough to account for the total observed VHE $\gamma$-ray 
emission. Nevertheless, a minor contribution is not excluded. Of the Wolf-Rayet \footnote{These stars
are in late phases of the evolution, exhibiting strong winds and large mass-loss rates.} 
(WR) star WR74, the Ant 
Nebula and the open stellar cluster Pismis 22, the latter, located at a distance of d$\sim$1 kpc, is 
the most promising candidate to explain the observed H.E.S.S. emission~\citep{Rowell2008}. Under the 
assumption that 10 stars of B-type (with 20\% efficiency) are associated to Pismis 22 
(see \citet{Rowell2008}), their stellar wind luminosities would be sufficient to power the observed 
VHE $\gamma$-ray emission. Additionally, with an estimated age of 40 Myr \citep{Piatti2000}, the system is old 
enough, that the most massive member stars could have already evolved into SNe, giving an extra 
contribution to the energy available for particle acceleration. Furthermore, XMM-Newton 
\citep{Rowell2008} and Suzaku \citep{Matsumoto2008} observations revealed an extended non-thermal 
X-ray emission towards the northern part of the source. Remarkably, the Suzaku `Src B' also shows 
an extended non-thermal nature and is centred on Pismis 22, similar to other open clusters connected 
to VHE $\gamma$-ray emission, like Cyg-OB2 \citep{Aharonian2005} or Westerlund 2 \citep{Aharonian2007}.

\section{Westerlund 2/HESS~J1023--575} \label{section:Wd2}
The prominent giant H II region RCW49 and its ionizing cluster Westerlund~2 are located towards 
the outer edge of the Carina arm in our Milky Way at a distance of 8 kpc, following \citet{Rauw2007}. 
Recently Chandra discovered $\sim$500 point 
sources in the vicinity of RCW49~\citep{Tsu04}, with $\sim$100 of them spatially coincident with 
the central open stellar cluster Westerlund~2~\citep{Tow04}. Mid-infrared measurements with Spitzer 
revealed still ongoing massive star formation in RCW49~\citep{Whi04}. The regions surrounding 
Westerlund 2 appear evacuated by stellar winds and radiation, in contrast to the dust distribution 
in fine filaments, knots, pillars, bubbles, and bow shocks throughout the rest of the H II 
complex~\citep{Chu04, Con04}. Radio continuum observations by ATCA at 1.38 and 2.38 GHz indicate 
two wind-blown shells in the core of RCW49~\citep{Whi97}, one surrounding the central stellar 
cluster, the other a Wolf-Rayet star WR20b. Westerlund 2, is centrally located within RCW49. 
It contains an extraordinary ensemble of hot and massive stars, presumably at least a dozen OB 
stars, and two remarkable Wolf-Rayet stars. The binary character of WR20a was only recently 
established. Both \citet{Rau04b} and \citet{Bon04} presented solutions for a circular orbit with 
a period of $\sim$3.7 days. The derived inclination angle of $(74.5\pm2.0)^\circ$ implies masses 
of $83.0 \pm 5.0$ and $82.0 \pm 5.0$\,M$_{\odot}$ for the primary and secondary component, respectively
\citep{Rau04b}. 
This puts the Wolf-Rayet binary WR20a as the most massive of all measured binary systems in our Galaxy. 

\begin{figure}
\centering{\plotone{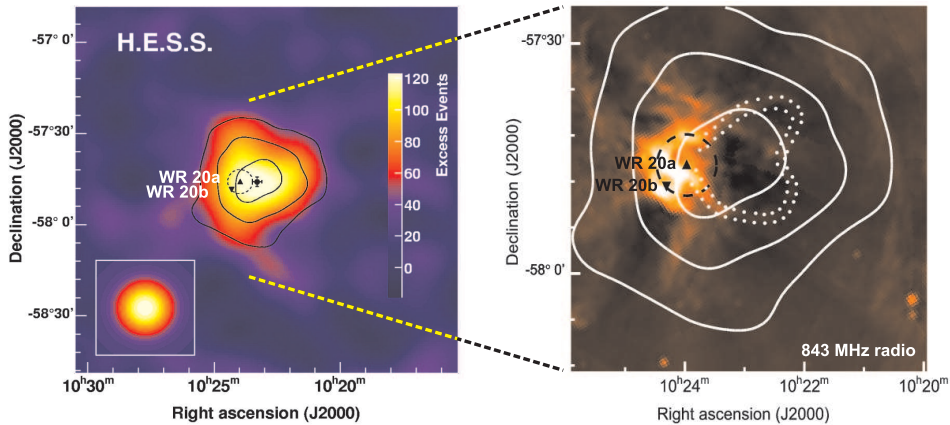}}
\caption{Left: H.E.S.S. $\gamma$-ray count map (oversampling radius of 0.12$^\circ$) of the 
Westerlund~2 region. The inlay in the lower left corner shows how a point-like source would 
have been seen by H.E.S.S. WR~20a and WR~20b are marked as filled triangles, and the stellar 
cluster Westerlund~2 is represented by a dashed circle. Right: Significance contours of the 
$\gamma$-ray source HESS~J1023--575 (5, 7 and 9$\sigma$ ), overlaid on a MOST 
radio image. The wind-blown bubble around WR~20a, and the blister to the west of it can be 
seen as depressions in the radio continuum. The blister is indicated by white dots as in 
\cite{Whi97}, and appears to be compatible in direction and location with HESS~J1023--575. 
(Figure taken from~\citet{Reimer2007}).
}\label{fig:Wd2}
\end{figure}

H.E.S.S. observed the region around Westerlund~2 in 2006 where a point source analysis on the 
nominal position of WR~20a resulted in a clear signal with a significance of 6.8$\sigma$. 
Further investigations revealed an extended excess with a peak significance exceeding 9$\sigma$ 
(Fig. \ref{fig:Wd2} left, \citep{Aharonian2007}). 
The source is clearly extended beyond the nominal extension of the point spread function 
(PSF), a fit of a Gaussian folded with the PSF of the H.E.S.S. instrument gives an extension of 
$0.18^\circ \pm 0.02^\circ$~\citep{Aharonian2007,Reimer2007}. The differential energy spectrum for 
VHE $\gamma$-rays inside the 85\% containment radius of 0.39$^\circ$ can be described by a power 
law (dN/dE$= \Phi_0 \cdot (\mbox{E}/1\,\mbox{TeV})^{- \Gamma}$) with a photon index of 
$\Gamma=2.53 \pm 0.16_{\mathrm{stat}} \pm 0.1_{\mathrm{syst}}$ and a normalization at 1\,TeV of 
$\Phi_0 = (4.50 \pm 0.56_{\mathrm{stat}} \pm 0.90_{\mathrm{syst}}) \times 10^{-12}$\,TeV$^{-1}$\,
cm$^{-2}$\,s$^{-1}$. The integral flux above 380 GeV is (1.3 $\pm$ 0.3) $\times 10^{-11}$\,cm$^{-2}$\,
s$^{-1}$. 

The detection of VHE $\gamma$-ray emission from the vicinity of Westerlund~2 can be explained by 
different acceleration mechanisms. One possibility is the conversion of kinetic energy available 
in the colliding-wind region of the binary system WR20a into the acceleration of particles to 
very high energies. Given the fact, that no significant flux variability could be detected in the 
data set and since the observed excess is extended with respect to the H.E.S.S. PSF, 
an association between WR20a and the VHE $\gamma$-ray emission is not striking. Alternatively, the 
emission could arise from the collective stellar wind effects in the Westerlund 2 cluster. As 
discussed before, particles are accelerated by multiple shocks and interact with the dense photon 
field of the stellar winds. However, the VHE $\gamma$-ray spectrum is expected to be similar that 
of SNRs at energies below 1~TeV. Particle acceleration by MHD motions (shocks, turbulence, etc.) 
of magnetized plasma produced by supersonic flows, which then penetrate into a dense medium may 
also be important under these conditions~\citep{Byk01}. The massive stellar winds of Westerlund 2 
could ensure sufficient particle injection into the turbulent plasma, feeding of magnetic turbulence 
with energy by wind-wind interactions of the massive star association, and the allocation of enough 
target material for the VHE $\gamma$-ray production. 

\section{Westerlund 1} \label{section:Wd1}
Motivated by the detection of VHE $\gamma$-ray emission towards Westerlund 2 and assuming similar 
particle acceleration mechanisms at work, Westerlund 1 (Wd~1) is an even more promising target for 
VHE $\gamma$-ray observations given that massive star content and distance are more favorable for 
detectable VHE $\gamma$-ray emission compared to Westerlund 2.

Westerlund~1 is the only known super star cluster in our galaxy. With a total mass of 
$\sim6\times 10^4~M_\odot$, it is currently the record holder in terms of its rich content of 
stars in the Wolf-Rayet phase. At least 24 WR stars are known of which $>$70\% are 
expected to be in binary systems \citep{Groh2006}. Additionally the existence of more than 
80 blue super-giants, 3 red super-giants, one luminous blue variable and 6 (out of 12 in the 
whole galaxy) yellow hyper-giants is reported in several papers (summarized in~\citet{Muno2006b}).
The dissipated power in the form of kinetic energy in the wind of the WR stars alone is 
$L_\mathrm{W}\approx 10^{39}$\ergs. Given an age of $\sim$ 5 Myrs \citep{Crow2006}, the most massive 
stars in Wd~1 evolved into SNe. Assuming furthermore a SN rate of 10$^{-4}$ yr$^{-1}$, the available 
kinetic energy for particle acceleration would be quadrupled \citep{Muno2006b}.

X-ray observations with the Chandra satellite have revealed a magnetar candidate 
(AXP CXOU J1647--4552, $P=10.6$~s) with a massive progenitor which is associated with Wd~1
\citep{Muno2006a}. A detailed analysis also revealed extended emission (of the order of arc minutes) 
which deviates from the typical thermal emission that can be seen from many other stellar 
associations. With increasing distance to the cluster core, the hard non-thermal emission dominates 
and line-emission disappears. The total luminosity observed in X-rays amounts to 
$L_x=3\times 10^{34}$\ergs \citep{Muno2006b}, representing just 10$^{-5}$ of the total mechanical 
luminosity in this system. The same authors discuss various possibilities to explain the `missing 
energy' in this powerful system, amongst others, the dissipation beyond the Chandra field-of-view 
by a large scale outflow and radiation at other wavelengths.
Recent measurements suggest a distance of the cluster of 4-5 kpc. Whereas the estimates by 
\citet{Brandner2005} and \citet{Crow2006} depend on the star content of the cluster and 
therefore suffer from extinction, \citet{Kothes2007} investigated the neutral environment of 
Wd~1 to find emission and absorption features in the HI data which could be connected to the stellar 
cluster. Interestingly they find two small expanding bubbles with dynamical ages of 
$\sim$0.6 Myrs and a large interstellar bubble with a size of 50 pc and a dynamical age of 
2.5 Myrs at a distance of $3.9\pm0.7$ kpc.

\begin{figure}[t!]
\centering{\plottwo{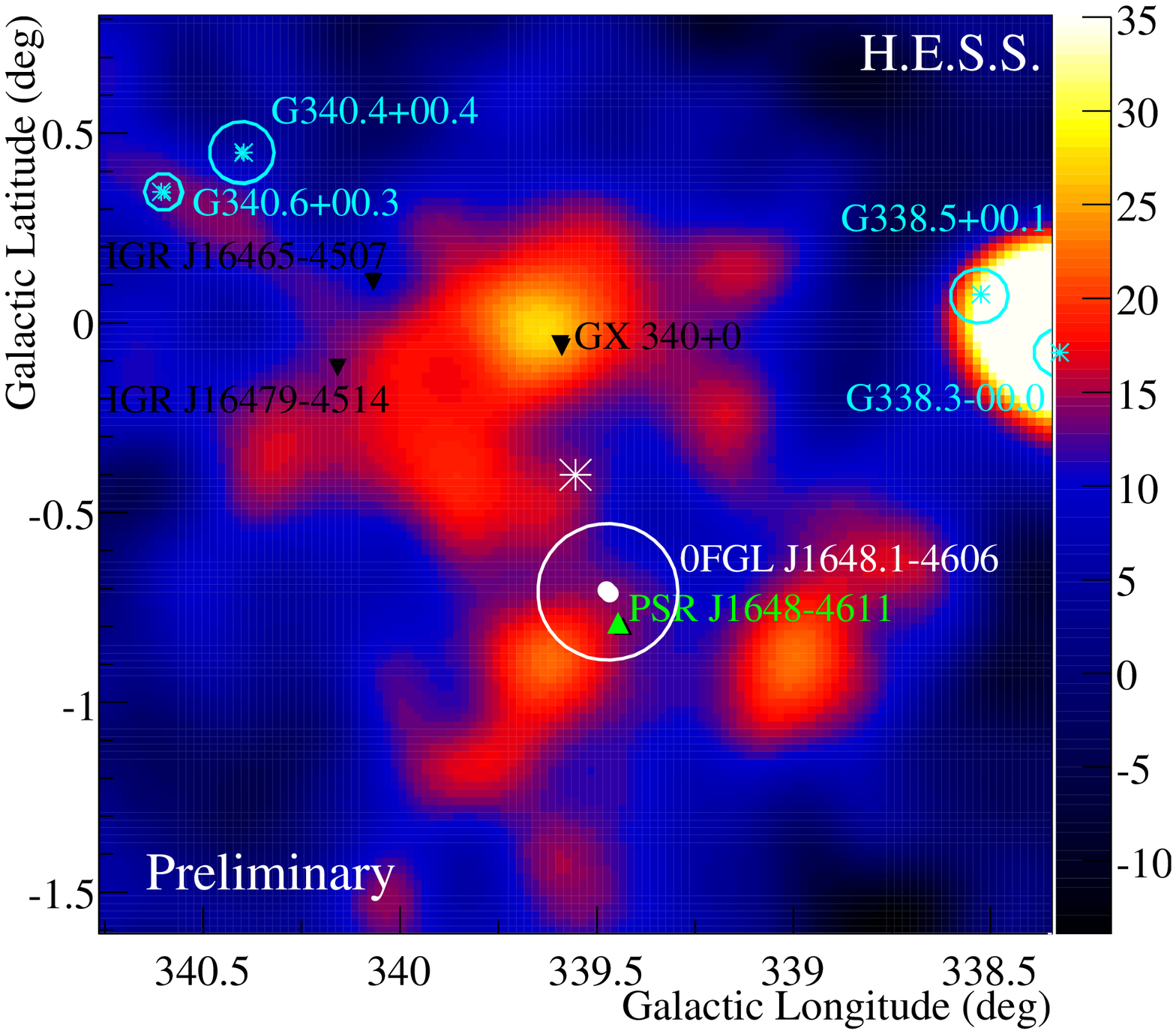}{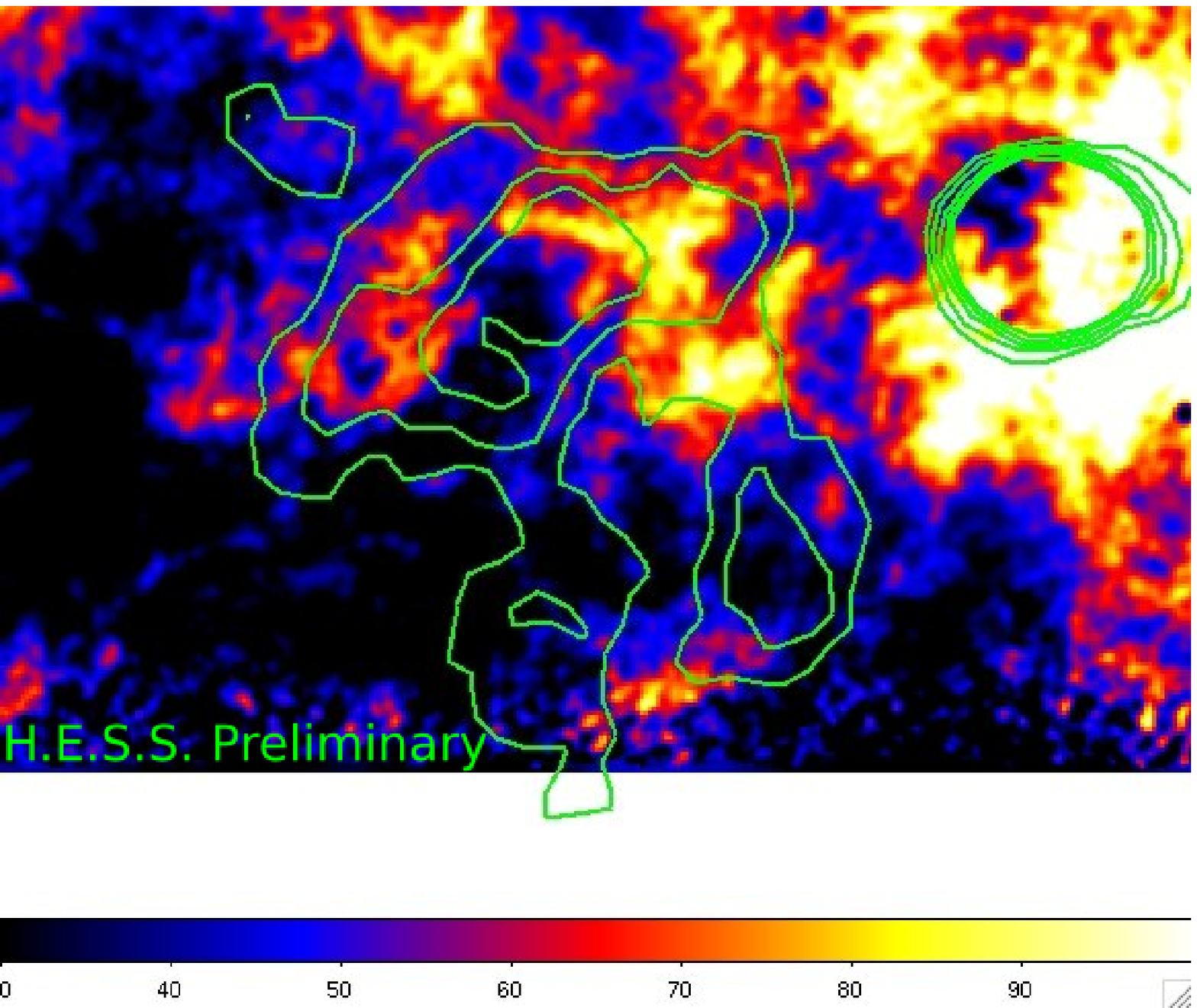}}
\caption{Left: Preliminary gaussian smoothed H.E.S.S. $\gamma$-ray count map ($\sigma =  0.13^\circ$) 
of the region around Wd~1 (indicated as white star), filled green and black triangles 
mark INTEGRAL sources and the pulsar PSR~J1648--4611, respectively. Cyan circles represent SNR 
candidates from Greens catalogue \citep{Green2004}. The white circle indicates the Fermi bright source 
OFGL~1648.1--4606 \citep{Fermi2009}. The bright region on the right is the known H.E.S.S. source 
HESS~J1640--465. Right: HI channel map of the Southern Galactic Plane Survey (SGPS) at a velocity of 
-55 km s$^{-1}$ \citep{McClure2005} corresponding to a distance of $\sim$4 kpc as obtained by 
\citet{Kothes2007}. Color scale from 20 K to 100 K, overlaid are the preliminary H.E.S.S. significance 
contours from 4 to 8 $\sigma$ in steps of 1$\sigma$ after integrating events within 0.22$^\circ$ radius 
(as done for the analysis of slightly extended sources compared to the H.E.S.S. PSF in the GPS).
}\label{fig:Wd1}
\end{figure}

H.E.S.S. observed the Westerlund 1 region from 2004 - 2007 during the GPS and in pointed observations 
for 14 hrs under good weather conditions (according to the Standard quality selection
\citep{Crab2006}) with the full array at zenith angles below 55$^\circ$. Another 22 hrs of good 
quality data was obtained in observations conducted between May - August 2008, resulting in a total 
data set of 34 hrs live-time \footnote{observation time, corrected for the dead-time of the 
system}. Preliminary and previously unpublished H.E.S.S. results of these observations reveal a 
spatially extended emission region of VHE $\gamma$-rays with a total significance of $> 15 \sigma$, 
within an integration radius of 1$^\circ$ around the Wd~1 position (85\% containment) 
(Fig. \ref{fig:Wd1}, left). 

The extension of the VHE $\gamma$-ray emission exceeds 2$^\circ$ in diameter, making it one of the largest 
structures observed in VHE $\gamma$-rays so far. A total of $>2300$ $\gamma$-rays are detected within the 
34 hours of livetime. The threshold of the analysis is at 680~GeV.

\begin{figure}
\centering{\plotone{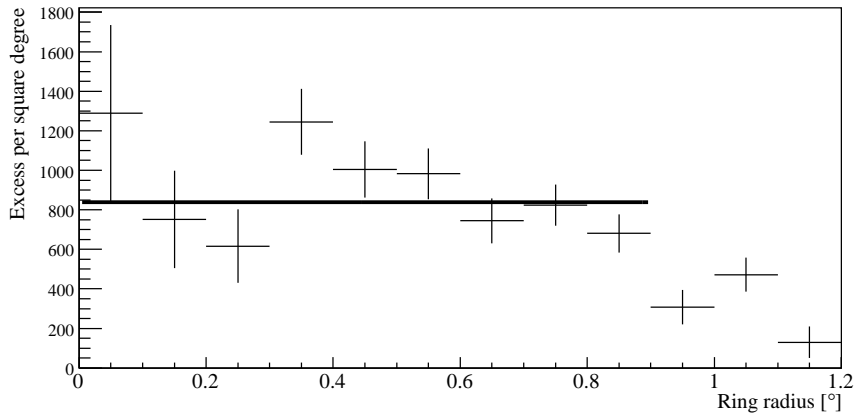}}
\caption{Preliminary radial profile of the uncorrelated $\gamma$-ray excess sky map 
starting at the Westerlund 1 position (white star in Fig. \ref{fig:Wd1}, left).
}\label{fig:Wd1_profile}
\end{figure}

Among the `established' classes of counterparts for VHE sources like SNRs and PWNe 
only PSR~J1648--4611 is coincident with the observed $\gamma$-ray emission. Given its 
$\dot{E} / d^2$ of $6.2\times10^{33} \mathrm{ergs}~\mathrm{s}^{-1}~\mathrm{kpc}^{-2}$ it can 
contribute to the VHE $\gamma$-ray emission. Recently, the LAT instrument on board of the Fermi 
satellite detected unpulsed emission from a region coincident with the pulsar position \citep{Fermi2009}. 
The Low Mass X-ray Binary (LMXB) GX~340+0 and other point-like counterparts are unlikely, given the 
rather extended nature of the VHE $\gamma$-ray emission. 
A radial profile of the excess counts per square degree in rings of 0.1$^\circ$ width starting 
from the Westerlund 1 position is shown in Fig. \ref{fig:Wd1_profile}. The VHE $\gamma$-ray 
emission from within the optical boundary of the stellar cluster (the innermost ring) is consistent with the 
overall picture of a rather flat emission out to a distance of 
$\sim$ 0.9$^\circ$ from the Wd~1 position (a fit of a constant yields a moderate $\chi^2$/ndf of 14.6/8). 
Fig. \ref{fig:Wd1}, right shows preliminary H.E.S.S. significance contours from 4 to 8 $\sigma$ in 
steps of 1 $\sigma$ after integrating events within 0.22$^\circ$ on top of the HI channel map at a 
velocity of -55 km s$^{-1}$, corresponding to the distance to Wd~1 as derived by \citet{Kothes2007}. The 
comparison may suggest a correlation between the VHE $\gamma$-ray emission and the HI 
data in some parts. Further spectral and morphological studies of the H.E.S.S. emission and observational 
data in radio and X-rays could help to shed some light on the origin of the TeV $\gamma$-ray emission and 
its possible connection to the super star cluster Westerlund 1.

\section{Discussion} \label{section:Discussion}

Massive stellar clusters provide sufficient kinetic energy in stellar winds, colliding wind binary 
systems and from collective wind and/or wind/SNe ejecta effects to accelerate particles to 
relativistic energies. After the detection of VHE $\gamma$-rays from the direction of the open stellar 
cluster Cyg-OB2 by the HEGRA Collaboration \citep{Aharonian2005}, H.E.S.S. discovered a source, 
coincident with the stellar cluster Westerund 2. Furthermore, one of the brightest unidentified H.E.S.S. 
sources, HESS~J1614--518, seems also be connected to a stellar cluster, Pismis 22. Westerlund~1 as the 
most massive stellar cluster known in our galaxy is a perfect target for VHE observations given its 
massive star content, age and binary fraction. The detection of extended emission from the vicinitiy of 
Westerlund~1 has been reported in this work. All these results suggest that multi-TeV particle 
acceleration may be linked to several massive stellar clusters.

\acknowledgements 
The author would like to thank the organizers for giving him the opportunity to 
give this invited talk at the HEPIMS workshop.
The support of the Namibian authorities and of the University of Namibia
in facilitating the construction and operation of H.E.S.S. is gratefully acknowledged,
as is the support by the German Ministry for Education
and Research (BMBF), the Max Planck Society, the French Ministry
for Research, the CNRS-IN2P3 and the Astroparticle Interdisciplinary
Programme of the CNRS, the U.K. Science and Technology Facilities
Council (STFC), the IPNP of the Charles University, the Polish Ministry
of Science and Higher Education, the South African Department of
Science and Technology and National Research Foundation, and by the
University of Namibia. We appreciate the excellent work of the technical
support staff in Berlin, Durham, Hamburg, Heidelberg, Palaiseau,
Paris, Saclay, and in Namibia in the construction and operation of the
equipment.


\end{document}